\documentclass{emulateapj}
\usepackage{color}
\usepackage{natbib}
\slugcomment{Submitted to Astrophysical Journal}
\shorttitle{Velocity Anisotropy -- Density Slope of DM Halos}
\shortauthors{Zait, Hoffman and Shlosman}

\def\gtorder{\mathrel{\raise.3ex\hbox{$>$}\mkern-14mu
    \lower0.6ex\hbox{$\sim$}}}
\def\ltorder{\mathrel{\raise.3ex\hbox{$<$}\mkern-14mu
    \lower0.6ex\hbox{$\sim$}}}

\def \via {{\it via\ }}
\def \ie {{\it i.e.,} }
\def \eg{{\it e.g.,}}

\newcommand {\rvir} {$R_{\rm vir}$}

\begin{document}

\title{Dark Matter Halos:  Velocity Anisotropy -- Density Slope Relation}

\author{
Amir Zait\altaffilmark{1},
Yehuda Hoffman\altaffilmark{1},
Isaac Shlosman\altaffilmark{2}
}
\altaffiltext{1}{
Racah Institute of Physics, Hebrew University; Jerusalem 91904, Israel
}
\altaffiltext{2}{
Department of Physics and Astronomy, 
University of Kentucky, 
Lexington, USA
}


\shorttitle{Velocity anisotropy -- density slope}
\shortauthors{Zait, Hoffman and Shlosman}


\begin{abstract}
Dark matter (DM) halos formed in CDM cosmologies seem to be characterized by  a  power law phase-
space density profile. The density of the DM halos is   often fitted by the   NFW profile but a better fit is provided by the   Sersic fitting formula.  
These relations are 
empirically derived from cosmological simulations of structure formation but have not yet been 
explained on a first principle basis. Here we solve   the Jeans equation under the
assumption of a spherical DM halo in  dynamical equilibrium, that obeys a power law phase space density and either the NFW-like or the Sersic density profile.
We then calculate the velocity anisotropy, $\beta(r)$, analytically. Our main result is that 
for the NFW-like  profile the $\beta ~ - ~ \gamma$ relation is not a linear one (where $\gamma$ 
is the 
logarithmic derivative of the density $\rho[r]$). The shape of $\beta(r)$ depends mostly on the 
ratio of the gravitational to kinetic energy within the NFW scale radius $R_{\rm s}$. 
For the Sersic profile   a linear  $\beta ~ - ~ \gamma$ relation is recovered, and in particular 
for  the Sersic index of $n\approx 6.0$ case the linear fit of Hansen \& Moore is reproduced. 
Our main result is that the phase-space density power law, the Sersic density form and the 
linear $\beta ~ - ~ \gamma$ dependence constitute a consistent set of relations which obey the spherical Jeans equation and as such provide the framework for the dynamical modeling of DM 
halos.
\end{abstract}


\section{Introduction}
\label{sec:intro}

In the standard cosmological model of structure formation the luminous matter is embedded 
in extended dark matter (DM) halos. The large scale structure emerges out of the primordial 
perturbation field \via\ gravitational instability. The model further assumes the DM  to be 
made of weakly interacting particles and its dynamics to be collisionless and therefore 
dissipationless (\citeauthor{pad93} \citeyear{pad93} for a review).  The problem of the 
dynamics of DM halos can be formulated as the classical gravitational $N$-body problem, 
subject to the assumption of cosmological initial and boundary conditions. As such the 
problem can be very easily formulated, yet it defies any rigorous analytical treatment. 
The spherical top-hat model provides the main analytical tool for shading light on the 
problem, but its scope of validity is rather limited \citep{gg72}. The model can be 
extended to accommodate shell crossing  \citep{gunn77} and cosmological initial conditions 
\citep{hs85} resulting in the secondary infall model.

In the absence of a rigorous analytical theory the study of the evolution DM halos relies  
heavily on numerical $N$-body simulations. The advent of CPU power and improved numerical 
algorithm have led to a general consensus about the basic properties of DM halo, such as 
the spherically-averaged density profile, the spin and shape of halos. One of the pillars 
of the phenomenology of DM halos is the so-called NFW density profile (\citeauthor{nfw96}, \citeyear{nfw97}),
\begin{equation}
\label{eq:NFW}
\rho_{\rm NFW}(r) = { 4 \rho_{\rm s} R_{\rm s}^3\over r (R_{\rm s} + r)^2}.
\end{equation}
It has been argued the NFW two-parameters fitting formula provides a good approximation 
for DM halos found in a wide range of mass scales and cosmological models, and hence can 
be considered universal. Subsequent numerical simulations have basically confirmed the 
functional form of the NFW profile but some controversy concerning the asymptotic slopes 
of the profile at small and large radii persists (\eg\  \citeauthor{moore98}  
\citeyear{moore98},  \citeauthor{kly01} \citeyear{kly01}, 
\citeauthor{jing00} \citeyear{jing00}). Here we refer to these generalizations of the 
NFW fitting formula as the gNFW profile.

A second basic phenomenological finding is the so-called phase-space density (PSD) profile 
\citep{tay01}. These authors defined the PSD profile by
\begin{equation}
\label{eq:phase-space}
Q(r)=\frac{\rho}{\sigma_{\rm r}^3},
\end{equation}
where $\sigma_{\rm r}$ is the radial velocity dispersion. The basic finding of \citet{tay01} is 
that the PSD profile follows a power law of the form
\begin{equation}
\label{eq:Q(r)}
Q(r) \propto r^{-\alpha}.
\end{equation}
\citet{tay01} found $\alpha\approx1.875$, and more recent studies find $\alpha=1.92 \pm 
0.01$ \citep{DM05} and $1.94 \pm 0.01$ \citep{HRSH07}.  Again, the power law behavior of 
$Q(r)$  is found over a wide range of mass scales and different cosmological models, 
suggesting a universal property of DM halos.

It has been recently suggested that a universal density slope- velocity anisotropy relation 
exists for relaxed DM halos \citep{HM06}. These author found a linear relation between the 
velocity anisotropy parameter $\beta$ and the density slope $\gamma$ ($<0$),
\begin{equation}
\label{eq:linear}
\beta = - 0.2(\gamma + 0.8),
\end{equation} 
where
\begin{equation}
\label{eq:beta}
\beta(r)=1-\frac{\sigma{_\theta^2}+\sigma{_\phi^2}}{2\sigma{_{\rm r}^2}}
\end{equation} 
and  
\begin{equation}
\label{eq:gamma}
\gamma={ r\over \rho} { d\  \rho \over  d\  r},
\end{equation} 
where $\sigma_\theta$ and $\sigma_\phi$  are the dispersions of the two transversal 
velocity components. Again, the linear relation has been found for a variety of halos 
obtained in various simulations and setups, suggesting a universal nature of the linear 
relation.

The phenomenological gNFW density profile, the PSD power law and the linear 
$\beta-\gamma$ relation are supposed to provide us clues about DM halos. The universal 
nature of these 
relations suggests that they hold over a broad range of scales and models, at least 
within the parameter space of the  Cold Dark Matter (CDM)-like cosmogonies. The equilibrium 
structure of collisionless self-gravitating systems obey the Jeans equation (\eg\ \citeauthor{BT}  
\citeyear{BT}), 
which relates the velocity second moments and the density field. Under the assumption of a 
spherical symmetry, the Jeans equation relates $\rho(r)$, $\sigma_{\rm r}(r)$ and $\beta(r)$, and 
can be rewritten so as  to relate $\rho(r)$, $Q(r)$ and $\beta(r)$. Lacking a fundamental 
theory that can predict even one of the above relations we are motivated to study the 
internal consistency of the three conditions and to find out whether one of these can be 
found to depend on the other two. Given the large scatter around the linear $\beta-\gamma$  
relation (\citeauthor{HM06} \citeyear{HM06},
\citeauthor{HS06} \citeyear{HS06}) we suspect that the $\beta -  \gamma$ relation is the 
'weakest' and less certain among the three relations. We therefore assume that the density  
of spherical halos follows an exact gNFW  profile and that the PSD profile is a power law 
and solve the Jeans equation to obtain the $\beta$ profile. This is to be compared with the 
linear $\beta - \gamma$ relation.

It has been recently suggested that the so-called  Sersic, also known as the    \cite{ein65}, profile  provides a better fit to the density profile \citep{mer05,prad06,gao07}.   Given the very different functional form of the Einaso profile it will be considered here as an alternative to the NFW-like family of profiles and the associated  $\beta(r)$ will be calculated. This will enable us to check the sensitivity of $\beta(r)$ to the assumed density fitting formula.

The structure of the paper is as follows. The Jeans equation is solved in 
\S \ref{sec:Jeans} and the solutions of the $\beta$ profile are given in \S \ref{sec:res} 
and discussed in \S \ref{sec:disc}.

\section{The Jeans equation}
\label{sec:Jeans}

The following model of DM halo is assumed here: the halos are in dynamical equilibrium; it 
is spherically symmetric; the halo density profile is given by the gNFW profile; and the 
PSD profile follows a power law. Such a halo should obey the Jeans equation (e.g., 
Binney \& Tremaine 1987),
\begin{equation}
\label {eq:jeans}
\frac{d\sigma{_{\rm r}^2}\rho}{dr}+\frac{2\beta\left(r\right)}{r}\sigma{_{\rm r}^2}\rho =
-\rho\left(r\right)\frac{GM\left(r\right)}{r^2},
\end{equation}
where $M(r)$ is the total mass enclosed within a spherical shell of radius $r$.

The PSD profile is expressed here in terms of the NFW scale radius, $R_s$ (Eq. \ref{eq:gnfw} 
below), the density, $\rho_{\rm s}$. and the radial velocity dispersion, $\sigma_{\rm r,s}$, 
evaluated at $R_{\rm s}$. Namely, 
\begin{equation}
\label{eq:phase-space}
Q(r) = {\rho_{\rm s}\over \sigma_{\rm r,s}^3} \bigg({r\over R_{\rm s}}\bigg)^{-\alpha}.
\end{equation}
The $(r/R_{\rm s})^{-\alpha}$ scaling is found to persist throughout  the entire evolution of 
individual DM halos along the main branch of their merger tree \citep{HRSH07}.

Rescaling the radial coordinate to $x=r/R_{\rm s}$ the Jeans equation is rewritten as:
\begin{eqnarray}
\left(\frac{\sigma_{\rm r,s}^{2}}{R_{\rm s}\rho_{\rm s}^{2/3}}\right)
\left[\frac{d}{dx}\left(\rho^{5/3}x^{2\alpha/3}\right)+
     \frac{2\beta\left(x\right)}{x}\rho^{5/3}x^{2\alpha/3}\right]  =  
       \nonumber\\
        -\rho\left(x\right)\frac{GM\left(x\right)}{\left(R_{\rm s} x\right)^{2}}.
\end{eqnarray}
Isolating $\beta(x)$ yields:
\begin{equation}
\label{eq:beta}
\beta(r) =  
-\frac{1}{2}\left[\frac{5x\rho'}{3\rho}+\frac{2\alpha}{3}+
    \frac{\rho_{\rm s}^{2/3}}{\sigma_{\rm r,s}^{2}x^{2\alpha/3}\rho^{2/3}}
       \frac{GM\left(x\right)}{R_{\rm s} x}\right]
\end{equation}	
Here the prime denotes a derivative with respect to $x$.

The density is assumed to follow a gNFW profile, using the functional form proposed by 
\citet{Zhao96}, namely
\begin{equation}
\label{eq:rho}
\rho(r) = \rho_{\rm s} \rho_{\rm gNFW} \bigg({r \over R_{\rm s}}\bigg),
\end{equation}
and
\begin{equation}
\label{eq:gnfw}
\rho_{\rm gNFW}\left(x\right)=\frac{2^{\epsilon-\mu}}{{x^\mu}\left(1+x\right)^{\epsilon-\mu}}.
\end{equation}
The NFW profile corresponds to $\mu=1$ and $\epsilon=3$.

Given the PSD power law and the the gNFW profile the structure of a DM halo is determined by 
three parameters, \ie\ $R_{\rm s}$, $\rho_{\rm s}$ and $\sigma_{\rm r,s}$. In particular the velocity 
anisotropy profile is given by:
\begin{eqnarray}
\label{eq:betag}
\beta (x)  
      & =  & -\frac{\alpha}{3}+\frac{5\left(\mu+\epsilon x\right)}{6\left(1+x\right)} + 
          \nonumber\\
      &    & \eta  \frac{\pi 2^{1+(\epsilon-\mu)/3} x^{2-\mu/3}
                \left(1+x\right)^{2(\epsilon-\mu)/3}}{\left(\mu-3\right)x^{2\alpha/3} }
                     \times \nonumber\\
      &    & {_2F_1} \left(\epsilon-\mu, 3-\mu,4-\mu, -x\right),
\end{eqnarray}
where ${_2F_1}\left(a,b,c,x\right)$ is the Gauss Hypergeometric function and 
$\eta = G\rho_{\rm s} R_{\rm s}^2/\sigma_{\rm r,s}^2$ is the dimensionless constant that 
determines $\beta$. Note that $\eta$ is the square of the crossing-to-dynamical time ratio 
within $R_{\rm s}$, hence it scales {\it inversely} with the (virial) ratio of 
kinetic-to-gravitational energy, $T/W$, within this radius. The kinetic energy term, $T$,
is calculated here
taking into account only the radial dispersion velocities, $\sigma_{\rm r}^2$.

The NFW density profile yields a $\beta$ profile of the form
\begin{eqnarray}
\beta_{\rm NFW}(x)  & =  & -\frac{\alpha}{3}+\frac{5\left(1+3x\right)}{6\left(1+x\right)}- 
              \nonumber\\ 
                &     & \eta \frac{2^{5/3}\pi 
                  \left[{\rm ln}\left(1+x\right)-\frac{1}{1+1/x}\right]
                   \left(1+x\right)^{4/3}}{x^{(2\alpha+1)/3}}.
\end{eqnarray}
Hence $\beta(x)$ has a minimum around $x\sim 0.1$ and a maximum somewhere between $x\sim 1$ 
and 10, depending on the value of $\eta$.

\begin{figure*}[!t]
\epsscale{0.95}
\plotone{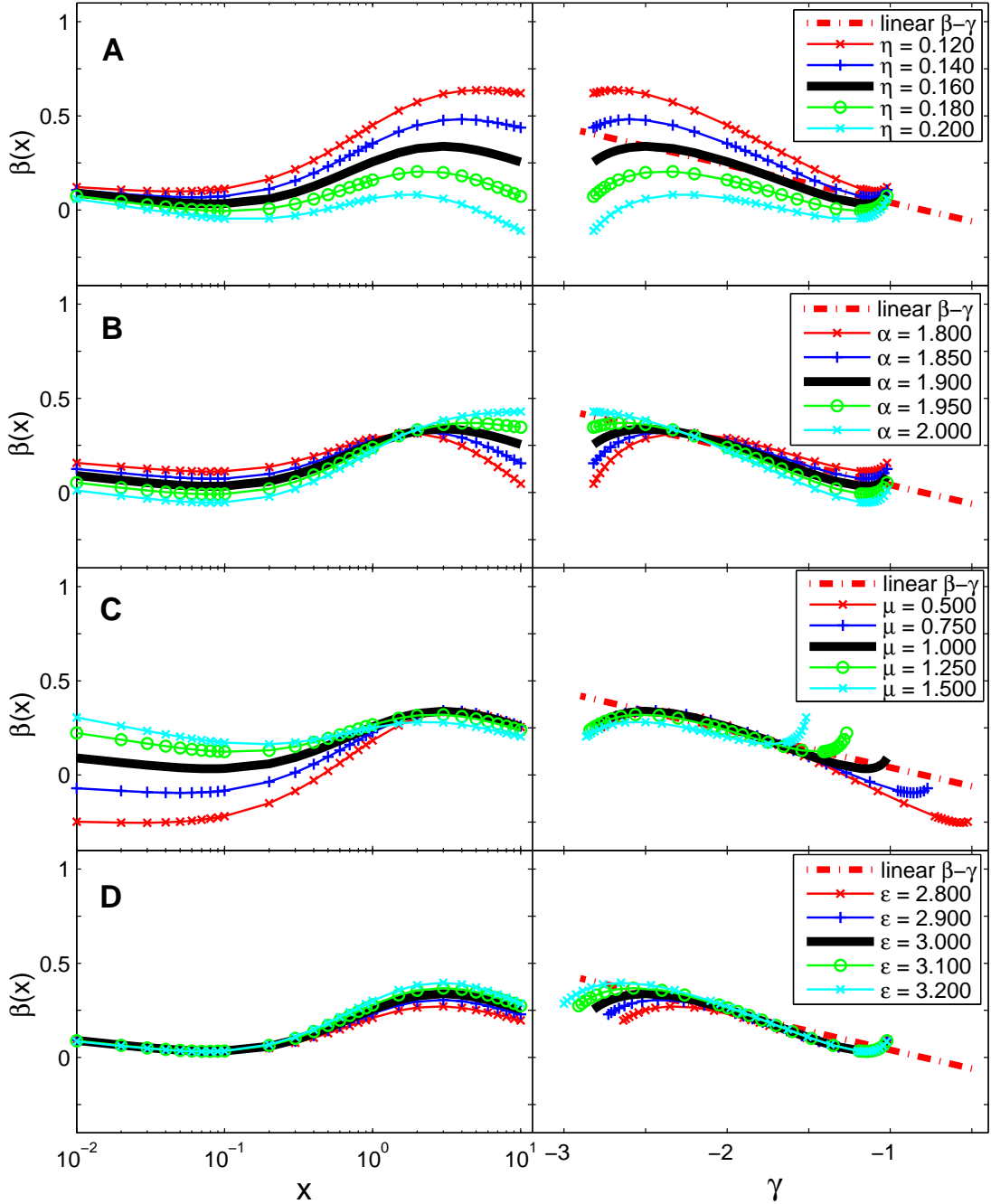}
\caption{
The dependence of the velocity anisotropy profile $\beta(x\equiv r/R_{\rm s})$ on the phase 
space density power law slope $\alpha$, the inverse virial parameter $\eta = G\rho_{\rm s} 
R_{\rm s}^2/\sigma_{\rm r,s}^2$ and the parameters of the generalized NFW density profile is 
presented. The left hand column displays 
the $\beta(x)$ and the right hand one the dependence of $\beta$ on the logarithmic derivative of 
the density profile $\gamma$. The Hansen \& Moore (2006) empirical fit to $\beta-\gamma$ is
plotted in the right frames. A nominal model is assumed ($\eta=0.16, ~\mu=1.0, ~\alpha=1.9$ and $\epsilon=3.0$) and the parameters are varied around that model.
}
\label{fig:beta}
\end{figure*}

The Sersic density profile is given by 
\begin{equation}
\label{eq:ein2}
\rho(r)= \rho_s \rho_{Ser}(r/R_{\rm s}) 
\end{equation}
where
\begin{equation}
\label{eq:ein3}
\rho_{Ser}(x)=\exp\Big[ -2 n (x^{1/n} -1)    \Big],
\end{equation}
and $n$ is the Sersic index \citep{prad06}. 
The Sersic   density profile yields a $\beta$ profile of the form
\begin{eqnarray}
\beta_{\rm Ser}(x)  & =  &  
{1 \over 3 \exp^{2/3}\big[-2n(x^{1/n}-1\big]}  \Big\{8^{-n} x^{-2\alpha/3}(n x^{1/n})^{-3n} \nonumber\\ 
&   &
\Big(8^n  \exp^{2/3}\big[-2n(x^{1/n}-1\big] x^{2\alpha/3} (n x^{1/n})^{3n}   \nonumber\\ 
&   &  (5 x^{1/n}-\alpha) - 2 \exp[2 n] \pi x^2 \eta \Gamma(1+3n)  \nonumber\\ 
&   &   + 6 \exp[2n] n \pi x^2 \eta \Gamma(3n,2 n x^{1/2})
\Big)
\Big\},
\label{eq:beta-ser}
\end{eqnarray}
where $\Gamma(x)$ and $\Gamma(a,x)$ are the usual $\Gamma$ and the incomplete $\Gamma$ functions.

\section{Results}
\label{sec:res}
\subsection{Generalized NFW Profile}
\label{sec:gnfw}

Under the assumption of a gNFW density profile and a power law PSD profile the $\beta$ profile 
is given by Eq. \ref{eq:betag}. Here we choose a base model and change the four halo parameters, 
one at the time, to check the dependence of $\beta$ (Fig. \ref{fig:beta}). The base model is an 
NFW density profile (\ie\ 
$\epsilon=3$ and a $\mu=1$ cusp), $\alpha=1.9$ (\citeauthor{DM05} \citeyear{DM05},
\citeauthor{HRSH07} \citeyear{HRSH07}). 
\citet{rd07} found that $T/W$ within $R_{\rm s}$ is roughly 2, or somewhat larger. 
Therefore, 
\begin{equation}
\label{eq:vir}
2 \approx {-2 T(<R_{\rm s}) \over W(<R_{\rm s}) } \approx {3 \over 4 \pi \eta}.
\end{equation}
For the base model we choose $\eta=0.15$. Fig. \ref{fig:beta} shows $\beta$ as a function of 
$x=r/R_{\rm s}$ (left column) and $\gamma$ (right column). The empirical 
$\beta=-0.2(\gamma + 0.8)$ relation  \citep{HM06} is plotted for reference. 

\begin{figure*}[!t]
\epsscale{0.95}
\plotone{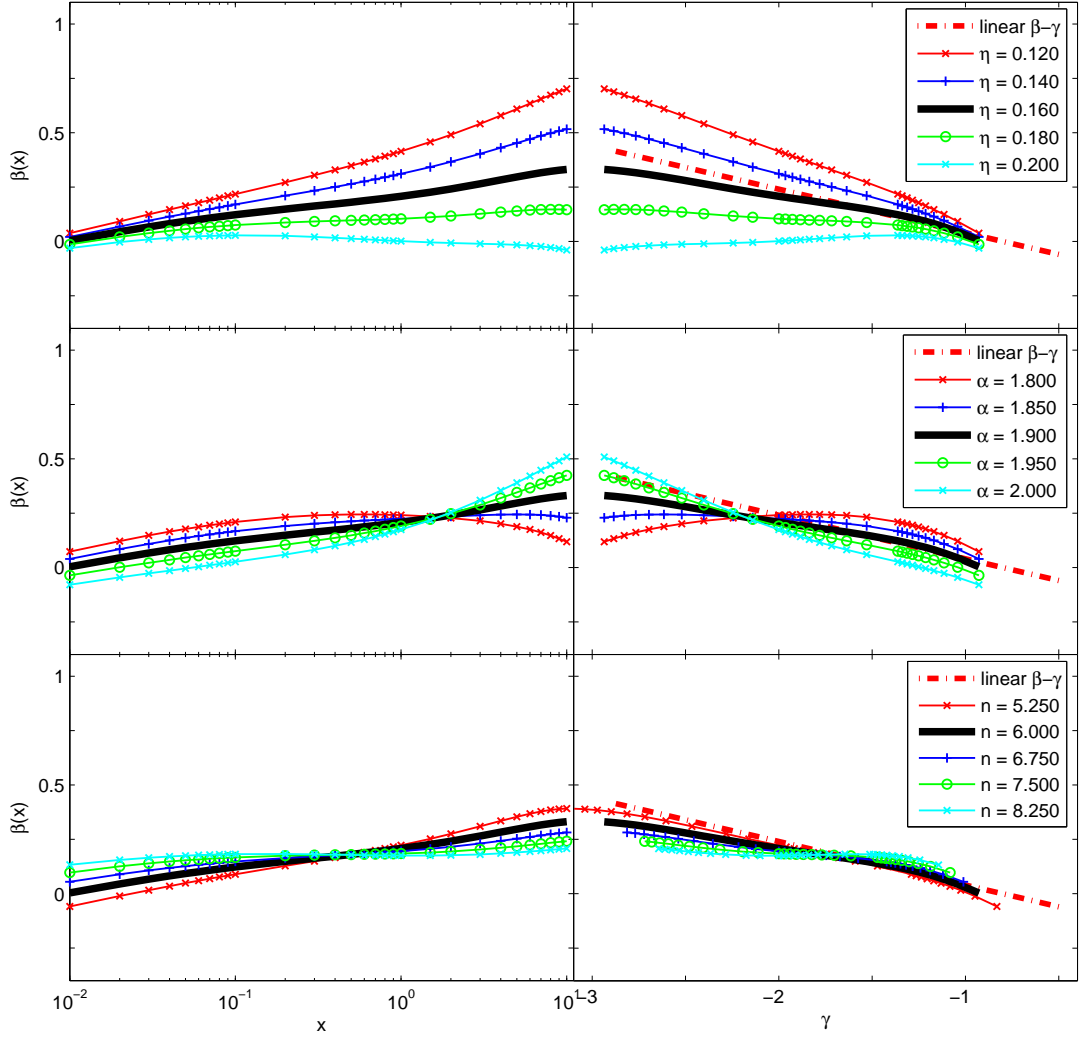}
\caption{
The dependence of the   $\beta(r)$ profile on $\alpha$, $\eta$   and the Sersic index $n$ is examined The structure of the figure is identical  to Fig. \ref{fig:beta}. The base Sersic model is taken here to have $1/n=0.16$  and the lower panel shows the variation of the Sersic index.
}
\label{fig:beta-sersic}
\end{figure*}

Over the range of parameters studied here $\beta(x)$ has local minimum around $\sim 0.1 R_{\rm s}$ and a local maximum at a few$\times R_s$. At the minimum the velocity dispersion is nearly isotropic 
(\ie\ $\beta \approx 0$. Moving outwards  to larger radii the velocity dispersion becomes more radial, as $\beta$ increases. However, beyond a few $R_s$ $\beta$ starts to decrease and the dispersion becomes more isotropic. It is interesting to note that on very small scales, below $0.1 R_s$, $\beta$ monotonically decreases with $r$. Its value at $r=0$ depends mostly on the slope of the cusp, $\mu$.  We find for $\mu=0.5, 1.0$ and  $1.5$ the asymptotic values of 
$\beta$ for $r\rightarrow 0$ are -0.21, 0.2 and 0.62, respectively. (Note that for $\mu=0.5$ $\beta$ reaches a minimum at $r \approx 0.02 R_s$ with $\beta \approx -0.25$.) 
The small scale behavior shows that NFW-like halos, obeying the phase-space density power law, can have steeper inner cusps by making their velocity dispersion more radial at the center. 
In other words, more radial orbits lead to a steeper inner density cusp.

Eq. \ref{eq:betag}, and consequently Fig. \ref{fig:beta}, displays the dependence of
$\beta$ on $\eta$  --- namely, an increase in $\eta$ leads to a decrease in $\beta$. The 
latter can be understood as following. Increase in $\eta$ means a decrease in $T/W$ ratio, or
equivalently an increase in $(v_{\rm c}/\sigma_{\rm r})^2$, where $v_{\rm c}$ is the circular 
velocity at radius $r$. Here two possibilities exist: (1) if $v_{\rm c}^2$ has increased at 
fixed $\sigma_{\rm r}^2$, the system responds by increasing the tangential dispersion 
velocities in order to remain in virial equilibrium and to be supported against the collapse. 
Note that while the overall angular momentum, $J$, of the system described by Eq. \ref{eq:jeans} 
is zero, the individual orbits have non-zero $J$ and are randomly oriented which is in fact 
the source of the tangential velocity dispersion. An increase in $v_{\rm c}^2$ 
then is translated into the increase in the tangential velocity dispersions in Eq.
\ref{eq:beta} and the associated decrease in $\beta$. (2) Alternatively, the growth in $\eta$ 
can come from the decrease in $\sigma_{\rm r}^2$, while  $v_{\rm c}^2$ is fixed. The latter one
requires that the tangential velocity dispersions stay unchanged, and, consequently, $\beta$
will decrease by the same token.

\subsection{The Sersic Profile}
\label{sec:ser}

The $\beta(r)$ profile of the Sersic fit  (Eq. \ref{eq:beta-ser})   is presented in Fig. \ref{fig:beta-sersic}. 
The base model is taken to be $\eta=0.16$, $\alpha=1.90$ and a Sersic index of $n=6.0$. The most striking feature of the $\beta$ profile in the Sersic case is its much simpler functional form compared with the gNFW case. For the nominal $\alpha=1.9$ case the $\beta - \gamma$ relation is very close to linear and $\beta$ grows monotonically with $x$ over the range of $ 10^{-2} < x < 10$.  Moreover, the  Sersic index of $n\approx 6$ very closely follows the  \citet{HM06}  linear relation. This is a surprising result. Both the gNFW and the Sersic profiles have been invoked as fitting parametric models of  the density profile in the DM halos, and as such they do not differ substantially. Yet, their resulting $\beta$ profiles show considerable qualitative differences.

\begin{figure}[!t]
\epsscale{0.95}
\plotone{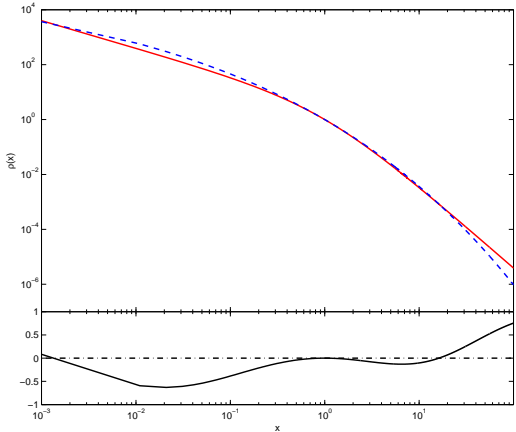}
\caption{
The NFW (red) and Sersic ($n=6$, blue) density profiles. The bottom panel shows the fractional difference between the two models. 
}
\label{fig:comp-rho}
\end{figure}
\begin{figure}[!t]
\epsscale{1.15}
\plotone{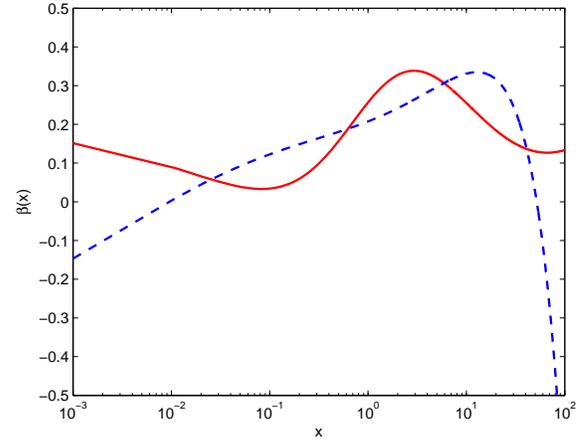}
\caption{
The NFW (red) and Sersic ($n=6$, blue) $\beta$ profiles. In both cases $\eta=0.16$ and $\alpha = 
1.9$ have been assumed. 
}
\label{fig:comp-beta}
\end{figure}

To gain further insight into the difference between the gNFW and the Sersic case we compare 
their corresponding  density  (Fig. \ref{fig:comp-rho}) and $\beta$ (Fig. \ref{fig:comp-beta}) 
profiles. The Sersic profile has been proved to be a good fit for the density profile of DM halos over the range of $10^{-2} < r /R_{\rm vir}  \lesssim (1-2)$, where \rvir\ is the virial radius 
\citep{prad06,gao07}. The NFW fitting has been done over a comparable range. Given that the 
Sersic and the gNFW profiles are just fitting formulae and are not derived from some  physical 
models, one should be cautious in using  them beyond the range over which the fit has been 
performed. To prove that point, Figs. \ref{fig:comp-rho} and \ref{fig:comp-beta} have been 
plotted over the range of $10^{-2} < x < 10^2$. The Sersic profile shows an unphysical drop at 
$x\approx 20$ where $\beta$ goes to minus infinity, namely the radial dispersion velocities 
vanish. This  behavior is shared by all the Sersic models considered here, of $5 \leqslant n 
\leqslant 8$. However, DM halos do not extend that far and this behavior should not affect the 
modeling of DM halos. The anisotropy parameter does not converge to a certain limit on very 
small scales, $x \ll 10^{-3}$ say, and the monotonic dependence on the radius is not guaranteed.

\subsection{Phase Space Density of the Total Velocity Dispersion}
\label{sec:qtot}

\begin{figure*}[!t]
\epsscale{0.95}
\plotone{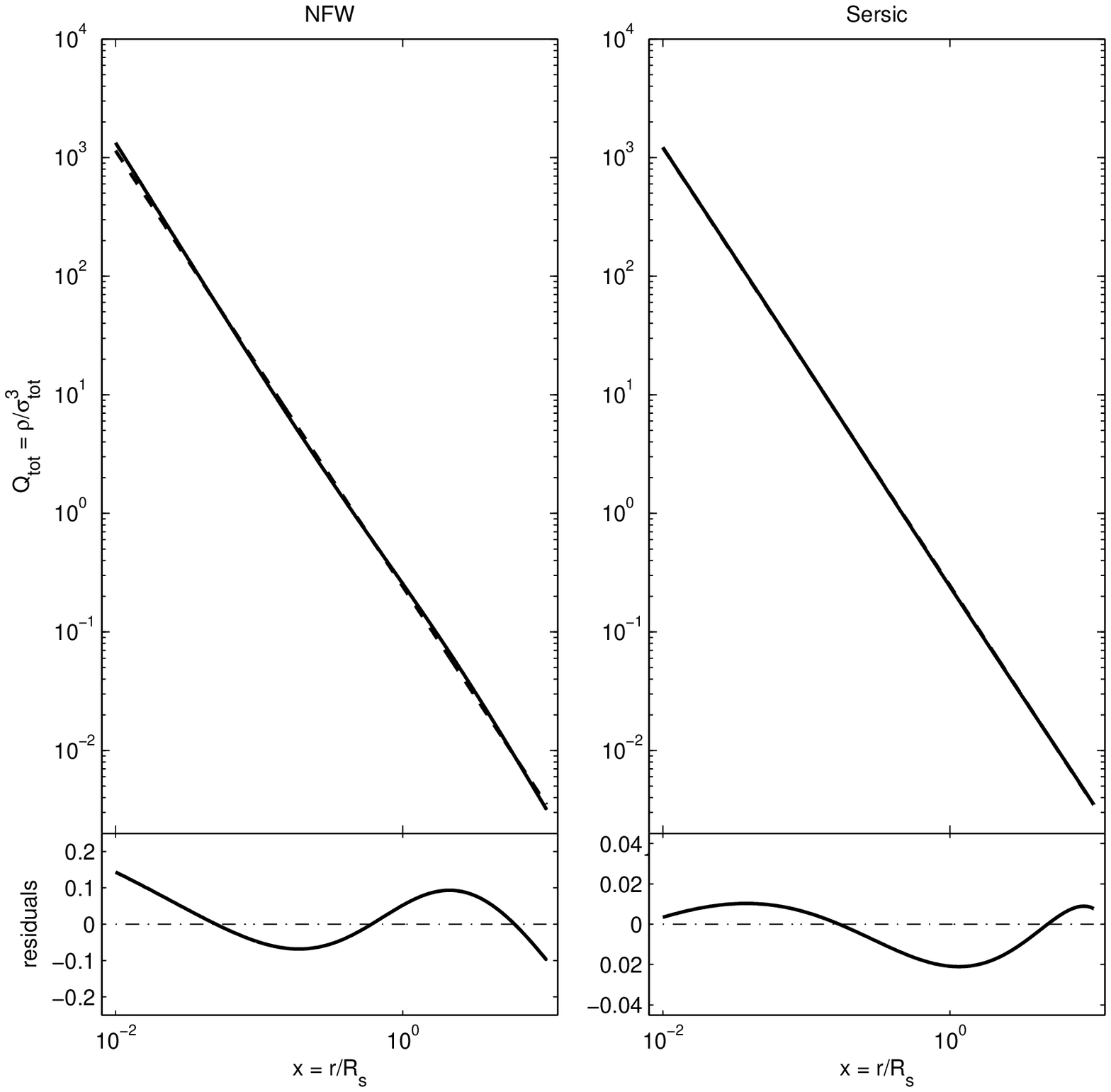}
\caption{
Given the power law $Q(r)$ profile and the calculated $\beta(r)$ the full velocity dispersion phase-space density profile,  $Q_{\rm tot}=\rho / \sigma{^2_{\rm tot}}$,  is 
easily evaluated. Here, $Q_{\rm tot}$ is evaluated for the   NFW density profile, 
($\eta=0.16$, left panel) and the Sersic model ($n=6$, right panel), assuming   $\alpha=1.9$ for both cases.
The total velocity PSD closely follows a power law, $Q_{\rm tot} \propto r^{-\alpha_{tot}}$, where $ \alpha_{\rm tot}=1.84$. The fractional residual from a power law is presented in the bottom panels.  Note  that in the NFW the amplitude of the residual is about $10$ percents and only about $1$ percent in the Sersic case.
}
\label{fig:qtot}
\end{figure*}

Given  the radial PSD and the calculated $\beta$ profile the PSD profile of the total velocity 
dispersion is easily evaluated. The total PSD is given by 
\begin{equation}
\label{eq:Qtot}
Q_{\rm tot}(r)=\frac{\rho}{\sigma{_{\rm tot}^3}} ,
\end{equation}
where $\sigma_{\rm tot}$ is the dispersion of the full three dimensional velocity field. This is 
related to the radial PSD $Q(r)$ by
\begin{equation}
\label{eq:Qtot-beta}
Q_{\rm tot}(r)=Q(r) [3 - 2 \beta(r)]^{-3/2} .
\end{equation}
Assuming the nominal parameters of $\eta=0.16$ and $\alpha=1.90$) the total PSD profile is calculated for the NFW and the Sersic ($n=6.0$) profiles (Fig. \ref{fig:qtot}).   In the two cases $Q_{\rm tot}(r)$ is very well fitted by an $r^{-\alpha_{\rm tot}}$  power law with $\alpha_{\rm tot}=1.847$ and $1.837$ for the Sersic and NFW cases, respectively.  The difference in the slope is insignificant and we conclude that   
$\alpha_{\rm tot}=1.84$. Yet, the fractional residual of the actual $Q_{\rm tot}(r)$ from the fitted power law are as large as twenty percents for the NFW case but they are of the order of one percent in the Sersic case. The $\alpha_{\rm tot}=1.84$   was found before by  \citet{DM05} and is in 
close agreement with the $1.82$ of 
\citet{fhgy07}.

\section{Discussion}
\label{sec:disc}

The present work has been primarily motivated by the quest for understanding  
the linear $\beta - \gamma$ dependence  \citep{HM06} and its relationship to the two  
so-called  universal relations, the PSD power law  \citep{tay01} and the gNFW and the Sersic density profiles, 
that characterize the structure of DM halos. Here we have focused on whether the 
linear $\beta - \gamma$ relation is implied by the other two relations, and is it consistent 
with them.

We find that for no choice of parameters  an NFW-like density profile can yield a velocity 
anisotropy,  $\beta(r)$, that is even in a rough agreement with the linear  $\beta - \gamma$. 
This stands in a sharp contrast with the Sersic fitting formula for the DM halos density profile. For a  Sersic profile index of   $n \approx 6.0$ one recovers quite faithfully the linear $\beta - \gamma$ relation. 
For that model $\beta(r)$ is a monotonically increasing function of the radius. For the nominal model we find $\beta(x =0.1)=0.12$ and $\beta(x=10)=0.33$. A by-product of the solution of the Jeans equation is that the PSD of the total velocity dispersion follows a power law, namely 
$Q_{\rm tot}(r) \propto r^{-\alpha_{\rm tot}}$, with $\alpha_{\rm tot}=1.84$, for both the gNFW and the Sersic density profiles. Yet, in the  fractional deviation in NFW-like case is of the order of twenty percents and in the Sersic case the deviation is of the order of one percent.

It has been realized in recent years that the Sersic profile provides a better fit to the density profile of DM halos  than the NFw  model 
\citep{mer05,mer06,prad06,gao07}. 
The present work substantiates and strengthens that fact and it strongly suggests that the the Sersic model should be used for the dynamical modeling of DM halos.

The main result of the papers is that the three pillars of the DM halos  phenomenology, namely the PSD power law, the Sersic density profile and the linear $\beta-\gamma$ relation, constitute a consistent set of relations that obey the Jeans equation. These relations provide  a theoretical framework for a consistent dynamical modeling of DM halos.  

A very different motivation for the calculation of the $\beta$ profile has been to provide a 
practical tool for modeling the mass distribution in clusters of galaxies from kinematic data. A 
powerful way of modeling  the clusters of galaxies is based on taking moments of the velocity 
distribution of clusters' galaxies and fitting them to the solutions of the Jeans equation, 
under the assumptions of the NFW density profile and $\beta= const.$ \citep{LO06,WL07}.  
The presently calculated $\beta$ profile 
certainly  provides a better approximation to the actual profile than the assumption of a 
constant value. We suggest that future analysis of the mass distribution of clusters will be 
based on the calculated profile of $\beta$.

We conclude the paper with a final note of caution.
The analysis presented here applies strictly to spherical DM halos in virial equilibrium. 
Inspection of the formation of DM halos in cosmological simulations of CDM-like cosmologies  
reveals an ongoing process where halos grow continuously by a slow accretion and by major 
mergers, alternating between phases of dynamical equilibrium and violent off-equilibrium. Therefore,
one expects the DM halos to show some deviations from a strict virial equilibrium and this might limit  
the validity of the solutions of the Jeans equation. 
DM halos are not isolated island structures in an otherwise unperturbed Friedmann universe. Halos are experiencing an ongoing smooth, and occasionally no so smooth, accretion and the boundary between the halo and outer universe is not easily defined. The stationary Jeans equation, on the other hand  is applicable to   isolated systems. It follows that even in stationary, seemingly relaxed systems  the Jeans equation may not be strictly obeyed.
A further complication arises 
from the deviation from sphericity of the DM halos. Cosmological simulations give rise to 
oblate and prolate ellipsoidal halos. This again might introduce a further deviation of the $\beta$ 
profile of the DM halos from the solutions of the Jeans equation. 
These issues should be further investigated by means of N-body simulations.

\acknowledgments

This research has been
supported by ISF-143/02 and the Sheinborn Foundation (to YH), and by
NASA/LTSA 5-13063, NASA/ATP NAG5-10823, HST/AR-10284 (to IS). 


\bibliography{/Users/hoffman/YH/papers/papers}

\end{document}